\documentstyle[12pt]{article}
\begin{document}

\begin{center}

{\large\bf Reply to Comment on ``Duality of}

{\large\bf  $x$ and $\psi$ in Quantum Mechanics"}

\vspace{1.cm}

 {\large Alon E. Faraggi$^{1}$ $\,$and$\,$ Marco Matone$^{2}$\\}
\vspace{.2in}
 {\it $^{1}$ Institute for Fundamental Theory, Department of Physics, \\
        University of Florida, Gainesville, FL 32611,
        USA\\
e-mail: faraggi@phys.ufl.edu\\}
\vspace{.025in}
{\it $^{2}$ Department of Physics ``G. Galilei'' -- Istituto 
                Nazionale di Fisica Nucleare\\
        University of Padova, Via Marzolo, 8 -- 35131 Padova, Italy\\
   e-mail: matone@padova.infn.it\\}

\end{center}

\vspace{0.50cm}

\noindent
The purpose of this note is to prevent possible confusion that may arise from
the misunderstanding in \cite{a} whose content is the derivation of Eq.(13) in
\cite{1} by direct differential calculus: which is precisely the same method we
used to derive it (it is in fact difficult to imagine any other possible
derivation). Because of the triviality of the derivation the details were not
given in \cite{1} and only the sentence ``In particular, by inverting the
Schr\"odinger equation we obtain'' was given before Eq.(13). On the other hand,
it is well known that the inversion of a differential equation is made by using
$\psi'(x)=1/x'(\psi)$ etc. (it would be bizarre to look for other methods). The
fact that Eq.(12) in \cite{1} precedes the explanation of its derivation
presumably caused the misunderstanding in \cite{a} where it is erroneously
stated that Eq.(13) was derived in \cite{1} from ``techniques inspired by field
theory duality". Actually, according to \cite{1}, substituting
$$
{\sqrt{2m}\over \hbar}x(\psi_E)={1\over2}{\psi_E}{\partial
{\cal F}_E\over \partial \psi_E}-{\cal F}_E,
$$
in the (inverted) Schr\"odinger equation 
${\hbar^2\over 2m}\partial_{\psi_E}^2x
=\psi_E(E-V(x))(\partial_{\psi_E}x)^3$ (=Eq.(13) in \cite{1}), one obtains
$4{\cal F}_E'''+(V(x)-E))\left({\cal F}_E'-\psi_E{\cal F}_E''\right)^3=0$. We
therefore clarify that no use of the prepotential ${\cal F}_E$ was made in the
derivation of (13), which is only one step in \cite{1} and from which (after
substituting (6) into (13)) the differential equation for ${\cal F}_E$ follows,
and not vice versa (which is, of course, well understood, e.g. \cite{vari}).
Finally, we note that in \cite{2} no reference to the pilot--wave interpretation
of quantum mechanics has been made. Referring in such a way to \cite{2} is the
interpretation of the authors of \cite{a}.

\end{document}